\documentclass[aps,twocolumn,amssymb,showpacs]{revtex4}
\usepackage{graphicx}

\begin{document}

\def\be{\begin{equation}}
\def\ee#1{\label{#1}\end{equation}}

\title{Cosmological models described 
by a mixture of van der Waals fluid and dark energy}
\author{G. M. Kremer}
\email{kremer@fisica.ufpr.br}
\affiliation{Departamento de F\'\i sica, Universidade Federal do Paran\'a,
Caixa Postal 19044, 81531-990 Curitiba, Brazil}

\begin{abstract}
The Universe  is modeled 
as a binary mixture whose constituents are described by a van der 
Waals fluid and by
a dark energy density. The dark energy density is considered
either as the quintessence or as the Chaplygin gas.  The irreversible 
processes 
concerning  the energy transfer between 
the van der Waals fluid and the gravitational field are taken into account. 
This model can 
simulate: (a) an inflationary period where 
the acceleration grows exponentially and the van der Waals
fluid behaves like an inflaton;
(b) an inflationary period 
where the acceleration is positive but it decreases and tends to zero 
whereas the energy
density of the van der Waals fluid decays;
(c) a decelerated period which corresponds to a matter dominated
period with a non-negative pressure; and (d) 
a present accelerated period where the dark energy
density outweighs the energy density of the van der Waals fluid.

\end{abstract}
\pacs{98.80.Cq}
\maketitle

\section{Introduction}

One of the main objectives of the cosmological models is the description 
of the 
different phases of the Universe concerning the time evolution 
of its acceleration field. The first epoch represents a rapid expansion of 
the Universe which is known as the  inflationary period. Most of
the theories that describe this period makes use of a scalar field which is 
related to a hypothetical particle, the so-called inflaton (see e.g. the 
works~\cite{Guth} and the references therein).
The second period refers to a past decelerated epoch where the energy 
densities of the radiation and matter fields outweigh the scalar field.
The present epoch is characterized by an accelerated Universe dominated
by a dark non-baryonic matter and a dark energy density.
The most common models for the dark energy density
make use of the equations of state of the 
quintessence~\cite{Stein} and of 
the  Chaplygin gas~\cite{KMP}.

Another model for the Universe was proposed recently by Capozziello 
and co-workers~\cite{CMF}   who used 
the equation of state of a van der Waals 
fluid in order to analyze the accelerated behavior 
of the Universe. The advantage of this model is that it can describe 
the transition 
from a scalar field dominated epoch to a matter field dominated period
without  introducing scalar fields.

The objective of the present work is to investigate a Universe described by 
the van der Waals equation of state. In section 2 we follow the 
works~\cite{CMF}
and model the Universe as a
non-dissipative one-component van der Waals fluid. We determine the time 
evolution of the
acceleration, energy density and pressure fields and show that this model
can simulate the two phases of the Universe beginning with  
an accelerated period and going into  a decelerated epoch.  
Although this model could describe the transition from an 
inflationary period to a matter dominated epoch  it could not 
simulate the present accelerated period of the Universe.  
For this end  we have modeled in section 3 the Universe
as a binary mixture of a van der Waals fluid and dark energy. The dark 
energy is
also regarded  either as the quintessence or as the Chaplygin gas. Furthermore,
we have consider the irreversible processes 
related to  the energy transfer between 
the van der Waals fluid and the gravitational field, since it is 
very questionable to get rid of the dissipative effects during the 
evolution of the Universe. Among other results,
it was shown that this model can 
simulate: (a) an inflationary period where 
the acceleration grows exponentially and the van der Waals
fluid behaves like an inflaton;
(b) an inflationary period 
where the acceleration is positive but it decreases and tends to zero 
whereas the energy
density of the van der Waals fluid decays;
(c) a decelerated period which corresponds to a matter dominated
period with a non-negative pressure; and (d) 
a present accelerated period where the dark energy
density outweighs the energy density of the van der Waals fluid.
Units have been chosen so that $c=1$ and $8\pi G/3=1$.

\section{Universe as a van der Waals fluid}

In this section we consider a model for a spatially flat, homogeneous
and isotropic  Universe where  
the energy-momentum tensor of the sources of the gravitational field 
is described by a perfect fluid with a van der Waals 
equation of state in the absence of  dissipative processes. 
The energy-momentum tensor $T^{\mu\nu}$ of the sources is given by 
\be
T^{\mu\nu}=(\rho_w+p_w)U^\mu U^\nu-p_wg^{\mu\nu},
\ee{1}
where $U^\mu$ (such that $U^\mu U_\mu=1$) is the four-velocity. The
pressure of the van der Waals fluid $p_w$ is related to its
energy density $\rho_w$ by (see, e.g.~\cite{Callen})
\be
p_w={8w_w\rho_w\over 3-\rho_w}-3\rho_w^2.
\ee{2}
In the above equation the pressure $p_w$ and  the energy density
$\rho_w$ are written in terms of dimensionless reduced variables and 
$w_w$ is a parameter connected with a reduced temperature. Here  $w_w$
will be  identified with the coefficient of the barotropic formula,
since for small values of $\rho_w$ we have $p_w\propto w_w\rho_w$.

By considering the Robertson-Walker metric and a comoving frame one
can  get from  the conservation law ${T^{\mu\nu}}_{;\nu}=0$
for the energy-momentum tensor
the following balance equation for the energy 
density
\be
\dot\rho_w+3H(\rho_w+p_w)=0.
\ee{3}
In the above equation $H=\dot a(t)/a(t)$ denotes the Hubble parameter,
$a(t)$ is the cosmic scale factor and the over-dot  refers 
to a differentiation with respect to a dimensionless time $t$.

The connection between the cosmic scale factor 
and the energy density is given by the  Friedmann equation 
\be
H^2=\rho_w,
\ee{4}
whose differentiation with respect to time
leads to 
the time evolution  of the cosmic scale factor 
\be
\dot H+{3\over 2}\left[H^2+{8w_wH^2\over 3-H^2}-3H^4\right]=0.
\ee{5}

The solution of the second-order differential equation  (\ref{5}) for $a(t)$ 
can be found by specifying initial values for $a(t)$ 
and $H(t)$ at $t=0$, for a given value of the parameter $w_w$. 
Here we have chosen the values $a(0)=1$ for the cosmic scale factor  and 
$H(0)=1$ for the Hubble parameter (by adjusting clocks) and in the following we
shall comment on the possible values for the parameter $w_w$.

In the interval $0\leq w_w<1/2$ the energy density grows with time,  
the pressure is always
negative whereas  the acceleration is  always positive. One can conclude
that this solution does not model properly the evolution of the Universe.

\begin{figure}\begin{center}
\includegraphics[width=6.4cm]{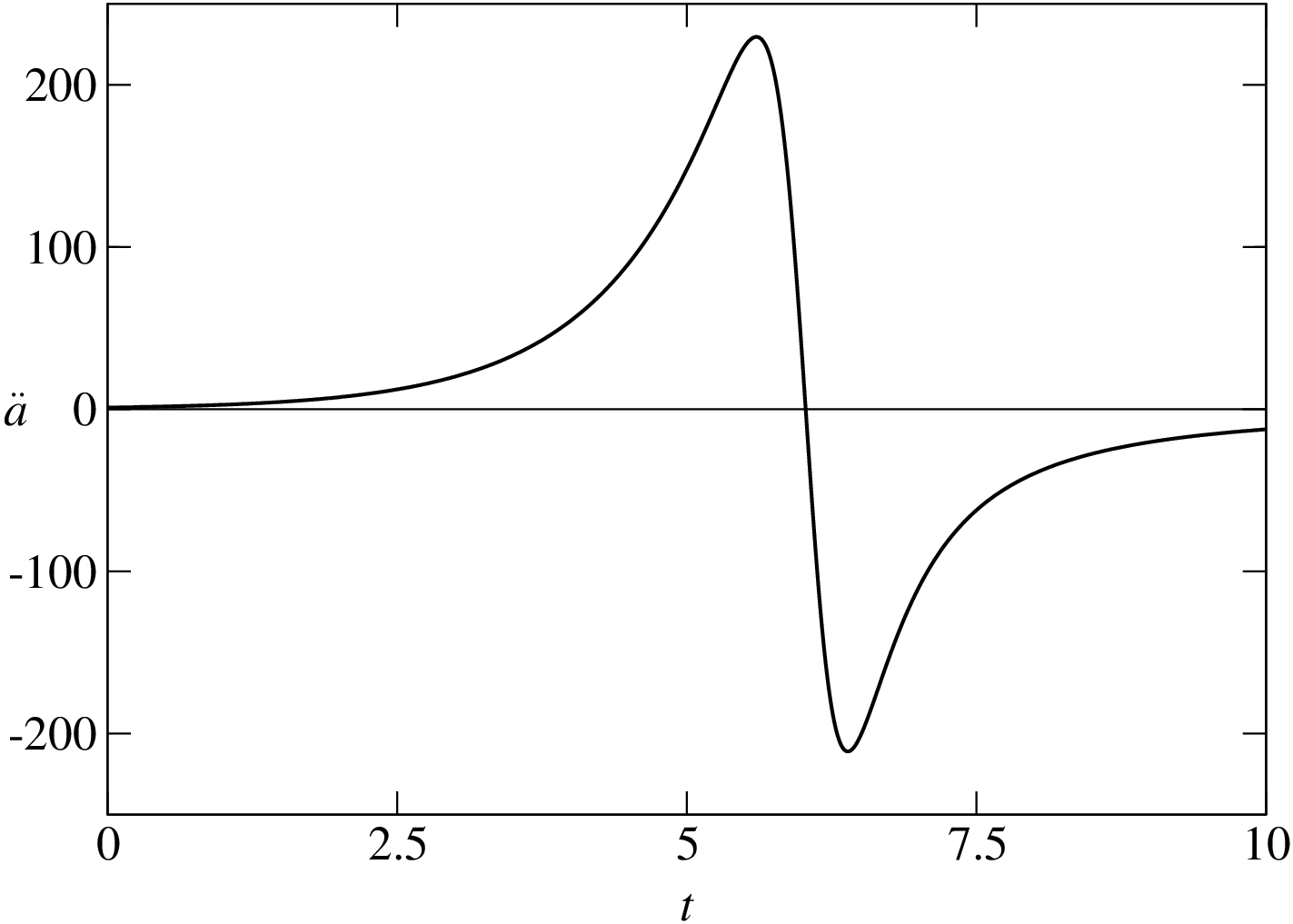}
\caption{Acceleration  vs time  for Universe 
as a van der Waals fluid.}
\end{center}\end{figure}

The case where $w_w=1/2$ is the most interesting   and   
in figures 1 and 2 we have plotted
the acceleration,
the energy density and the pressure as functions of the time. 
One can infer from figures 1 and 2 that the time interval between 
$0\leq t{^<_\sim} 5.6$ refers to an inflationary period since 
the acceleration grows exponentially and the van der Waals
fluid behaves like an inflaton with an equation of state $p_w=-\rho_w$. 
The time interval between
$5.6{^<_\sim}t{^<_\sim}6$ is also  another inflationary period 
since the acceleration is positive but it decreases and tends to zero 
whereas the energy
density decays and the pressure is always negative. For 
$6{^<_\sim}t{^<_\sim}6.4$ the 
acceleration is negative and attains its minimum value whereas 
the pressure is positive
and reaches its maximum value, hence this time interval can be 
interpreted as a matter dominated
period. The next time interval between $6.4{^<_\sim}t<\infty$ refers also 
to a matter dominated
period since the acceleration is always negative and the energy 
density and the pressure 
decay and tend to zero which corresponds to a dust dominated period.

\begin{figure}\begin{center}
\includegraphics[width=6.4cm]{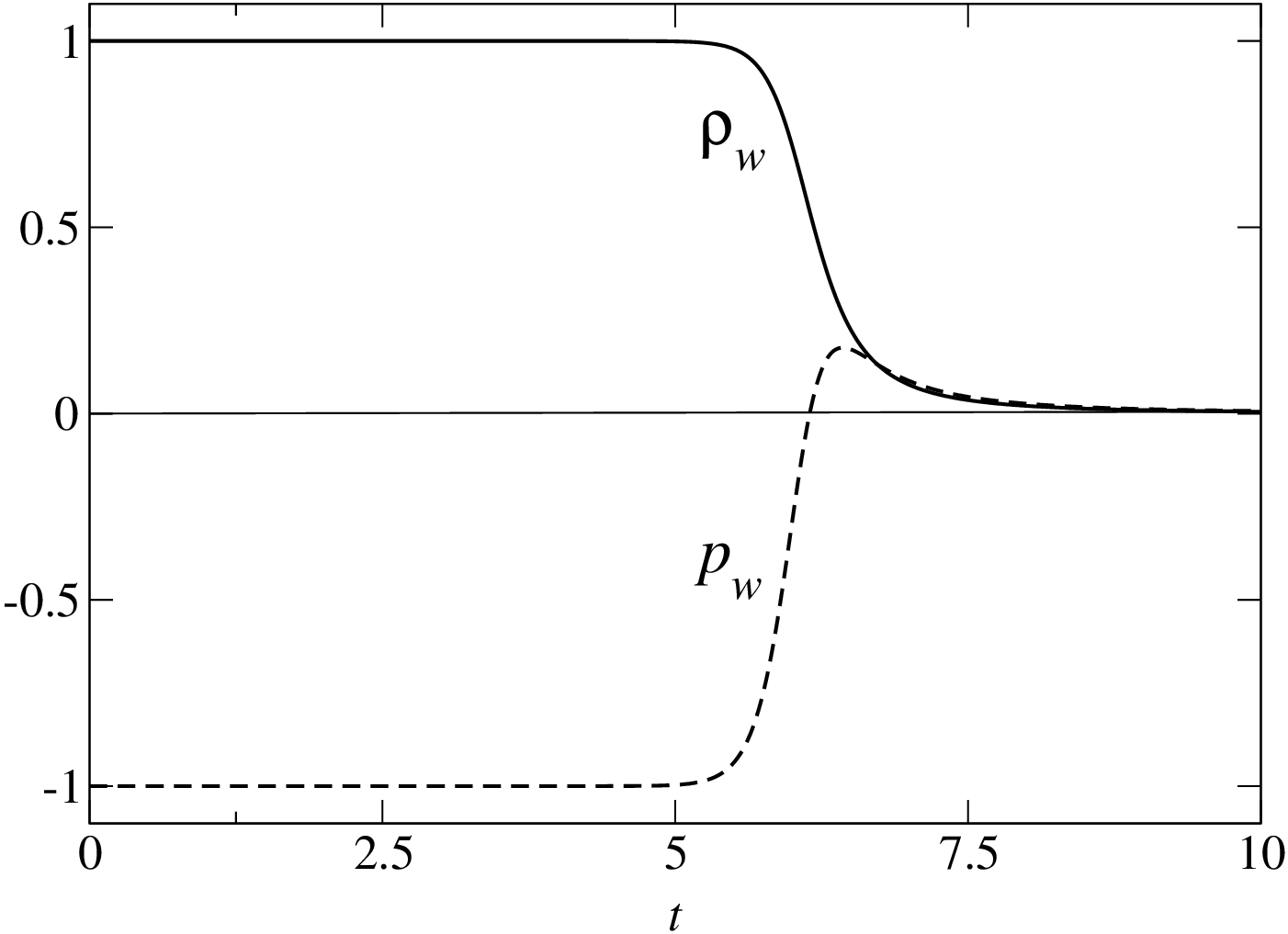}
\caption{Energy density (straight line) and pressure 
 (dashed line)  vs time for Universe as a van der Waals fluid.}
\end{center}\end{figure}

For  $1/2<w_w<2/3$ the energy density decays with time, 
the pressure grows from a negative value
to a maximum positive value and decays at large values of time 
while there exists a period of 
acceleration followed by a period of deceleration. This interval 
for $w_w$ could 
simulate  an inflationary period $(p_w<0)$, where the energy 
density of the inflaton decays,
followed by a  matter $(p_w\neq0)$ and dust $(p_w\rightarrow0)$ 
dominated periods.

The energy density and the pressure 
have the same behavior as the former interval for $2/3\leq w_w <3/4$  
but there exists 
only a period of deceleration and one can infer that    
this solution does not model properly the evolution of the Universe.

For $w_w \geq3/4$ the pressure is always 
positive whereas 
the acceleration is always negative and one can conclude that it 
refers only to a 
matter $(p_w\neq0)$ and dust $(p_w\rightarrow0)$ dominated periods.

Although the van der Waals fluid could describe the transition from an 
inflationary period to a matter dominated epoch  it could not 
simulate the present accelerated period of the Universe.  Moreover, in
the present model we have not consider the irreversible processes 
concerning the energy transfer between 
the van der Waals fluid and the gravitational field. This will be 
the subject of the next section.

\section{Universe as a binary mixture of  van der Waals fluid
and dark energy}

If  dissipative effects are taken into account and we
consider the Universe as a binary mixture of 
a van der Waals fluid with dark energy, the energy-momentum tensor reads
\be
T^{\mu\nu}=(\rho_w+\rho_d+p_w+p_d+\varpi)U^\mu U^\nu
-(p_w+p_d+\varpi)g^{\mu\nu}.
\ee{6}
Above $\rho_d$ is the dark energy density, 
$p_d$  its pressure  while $\varpi$ denotes 
the non-equilibrium pressure
which is connected with the irreversible 
processes (see e.g.~\cite{K2,Be}).

The evolution equation for the energy density of the mixture is now given by
\be
\dot\rho_w+\dot\rho_d+3H(\rho_w+\rho_d+p_w+p_d+\varpi)=0,
\ee{7}
while the Friedmann equation becomes
\be
H^2=\rho_w+\rho_d.
\ee{8}

The dark energy is supposed to interact only with itself and it is minimally 
coupled to the gravitational field, so that its energy density 
balance equation decouples
from that of the van der Waals fluid, and we obtain from  (\ref{7})
two balance equations
\be
\dot\rho_d+3H(\rho_d+p_d)=0,\qquad \dot\rho_w+3H(\rho_w+p_w)=-3H\varpi.
\ee{9}
The term $-3H\varpi$ on the right-hand side of (\ref{9})$_2$ can be 
interpreted 
as the energy density production rate of the van der Waals fluid. 
It is connected
with the  energy density production rate of the gravitational field by (see 
e.g.~\cite{K2}) 
\be
\dot\rho_G+3H(\rho_G-p_w-p_d)=3H\varpi,
\ee{10}
where $\rho_G$ is the energy density of the gravitational field. Hence, the 
non-equilibrium pressure is responsible for the irreversible transfer of the
energy density of the gravitational field to the van der Waals fluid.

Two models for the dark energy are considered here, namely the quintes\-sence
($\rho_d=\rho_X$) and the Chaplygin gas ($\rho_d=\rho_c$). The equations of 
state 
for these two models are given by 
\be
\cases{p_X=w_X\rho_X,\quad\hbox{with}\quad w_X<-1/3,\cr
p_c=-A/\rho_c,\,\quad\hbox{with}\quad A=\hbox{constant}>0.}
\ee{11}
For the  motivation of the  equations of state above one is referred to 
e.g.~\cite{Stein,KMP}
and the references therein.

The equations of state (\ref{11}) permit us to get from (\ref{9})$_1$ a 
relationship
between the energy density  of the quintessence and the cosmic scale factor 
as well as the corresponding one for the Chaplygin gas. In terms of 
dimensionless quantities these relationships read
\be
{\rho_X\over\rho_w^0}={\rho_X^0\over \rho_w^0}\left({1\over a}
\right)^{3(w_X+1)},\qquad 
{\rho_c\over\rho_w^0}={\rho_c^0\over \rho_w^0}{1\over \sqrt{1+\psi}}
\sqrt{1+{\psi\over a^6}},
\ee{12}
where $\rho_X^0/\rho_w^0$ and $\rho_c^0/\rho_w^0$ refer to the initial amount 
(at $t=0$ by adjusting clocks)
of the energy densities of the quintessence and of the Chaplygin gas with 
respect 
to the energy density of the van der Waals fluid, respectively  and $a$ is a 
dimensionless cosmic scale factor. The 
Chaplygin gas could interpolate a matter dominated
Universe (pressure-less fluid or dust) where $\rho_c\propto 1/a^3$
when  $\psi/a^6\gg1$   
 and a cosmological constant dominated Universe where $\rho_c=-p_c$ 
when $\psi/a^6\ll1$. Hence, the parameter $\psi$ -- which is related to 
the integration 
constant of the differential equation (\ref{9})$_1$
-- could be identified with 
the amount of the pressure-less fluid.

The time evolution of the cosmic scale factor (\ref{5}) for the mixture 
of a van der Waals fluid with quintessence becomes
$$
\dot H+{3\over 2}\left\{H^2+{8w_w[H^2-\rho_X^0/(\rho_w^0a^{3(w_X+1)})]\over 3-
[H^2-\rho_X^0/(\rho_w^0a^{3(w_X+1)})]}\right.
$$
$$\left.
-3\left[H^2-{\rho_X^0\over 
\rho_w^0}\left({1\over a}\right)^{3(w_X+1)}\right]^2
\right.
$$
\be\left.
+w_X{\rho_X^0\over 
\rho_w^0}\left({1\over a}\right)^{3(w_X+1)}+\varpi\right\}=0.
\ee{13}

Equation (\ref{13}) together with  an equation for the non-equilibrium 
pressure 
lead to a
complete determination of the time evolution of the cosmic scale factor
for a mixture 
of a van der Waals fluid with quintessence. 
Within the framework of extended (causal or second-order) thermodynamic 
theory  the evolution equation for the non-equilibrium pressure in a 
linearized theory reads~\footnote{For a derivation of (\ref{14})
from the relativistic Boltzmann equation see e.g. C. Cercignani  
and  G. M. Kremer, 
{\it The Relativistic Boltzmann
Equation: Theory and Applications}, (Birkh\"auser, Basel, 2002).}
\be
\varpi+\tau\dot\varpi=-3\eta H\quad\hbox{or}\quad\varpi+\alpha\dot\varpi
=-3\alpha H^3,
\ee{14}
by assuming that 
the coefficient of bulk viscosity $\eta$ and
the characteristic time $\tau$ are related to the energy 
density of the mixture $\rho=\rho_w+\rho_d$ (here $\rho_d=\rho_X$)
by $\eta=\alpha\rho$ with 
$\tau={\eta/ \rho}$ where $\alpha$ is a constant (see 
e.g.~\cite{K2}).

Once the cosmic scale factor is determined as a function of time, the
energy density of the quintessence can be calculated from
(\ref{12})$_1$,  while the energy density of the van der Waals
fluid  is obtained from
\be
{\rho_w\over \rho_w^0}=\left[H^2-{\rho_X^0\over 
\rho_w^0}\left({1\over a}\right)^{3(w_X+1)}\right].
\ee{14a}

For the mixture of a van der Waals fluid with a Chaplygin gas instead of 
(\ref{13}) we have
$$
\dot H+{3\over 2}\left\{H^2+{8w_w[H^2-\rho_c^0\sqrt{1+\psi/a^6}/(\rho_w^0
\sqrt{1+\psi})]\over 3-
[H^2-\rho_c^0\sqrt{1+\psi/a^6}/(\rho_w^0
\sqrt{1+\psi})]}\right.
$$
$$\left.
-3\left[H^2-{\rho_c^0\over \rho_w^0}{1\over\sqrt{1+\psi}}
\sqrt{1+{\psi\over a^6}}\right]^2
\right.
$$
\be\left.
-{\rho_c^0\over \rho_w^0}{1\over\sqrt{1+\psi}
\sqrt{1+{\psi/a^6}}}+\varpi\right\}=0.
\ee{15}
From the system of differential equations (\ref{14}) and (\ref{15}) it is
possible  to determine the 
time evolution of the cosmic scale factor
for a mixture 
of a van der Waals fluid with a Chaplygin gas. Moreover, the energy density 
of the Chaplygin gas follows from (\ref{12})$_2$ and the energy density 
of the van der Waals fluid can be determined from
\be
{\rho_w\over \rho_w^0}=\left[H^2-{\rho_c^0\over \rho_w^0}{1\over\sqrt{1+\psi}}
\sqrt{1+{\psi\over a^6}}\right].
\ee{16a}

The system of differential equations (\ref{13}) and (\ref{14})
for the mixture of a van der Waals fluid and quintessence
and the corresponding one (\ref{15}) and (\ref{14})
for the mixture of a van der Waals
fluid and a Chaplygin gas have been solved by considering the 
initial conditions: $a(0)=1$ for the cosmic scale factor,
$H(0)=1$ for the Hubble parameter and $\dot\varpi(0)=0$
for the non-equilibrium pressure. In order to have a complete determination
of the time evolution of the cosmic scale factor and of the energy densities
there still remains much freedom, since each system of differential
equations does depend on four parameters, namely: $w_w$, $\rho_X^0/ \rho_w^0$,
$w_X$, and $\alpha$ for the mixture with the quintessence and
$w_w$, $\rho_c^0/ \rho_w^0$, $\psi$, and $\alpha$ for the mixture
with the Chaplygin gas. These parameters  can be interpreted as follows:
(a)  $w_w$ is connected with the inflaton, since the increase of 
$w_w$ leads to a less pronounced inflationary period (see previous section);
(b) $\rho_X^0/ \rho_w^0$ and $\rho_c^0/ \rho_w^0$ are related to the initial
amount of the energy density of the quintessence and of the Chaplygin gas
with respect to the energy density of the van der Waals fluid; (c) $w_X$
takes into account the strength of the negative pressure of  the quintessence;
(d) $\psi$ -- as was previously commented -- is related to the amount of the 
pressure-less
fluid in the Chaplygin equation of state  and (e) $\alpha$ is connected with
the importance of the irreversible processes in the evolution of the Universe.

\begin{figure}\begin{center}
\includegraphics[width=6.4cm]{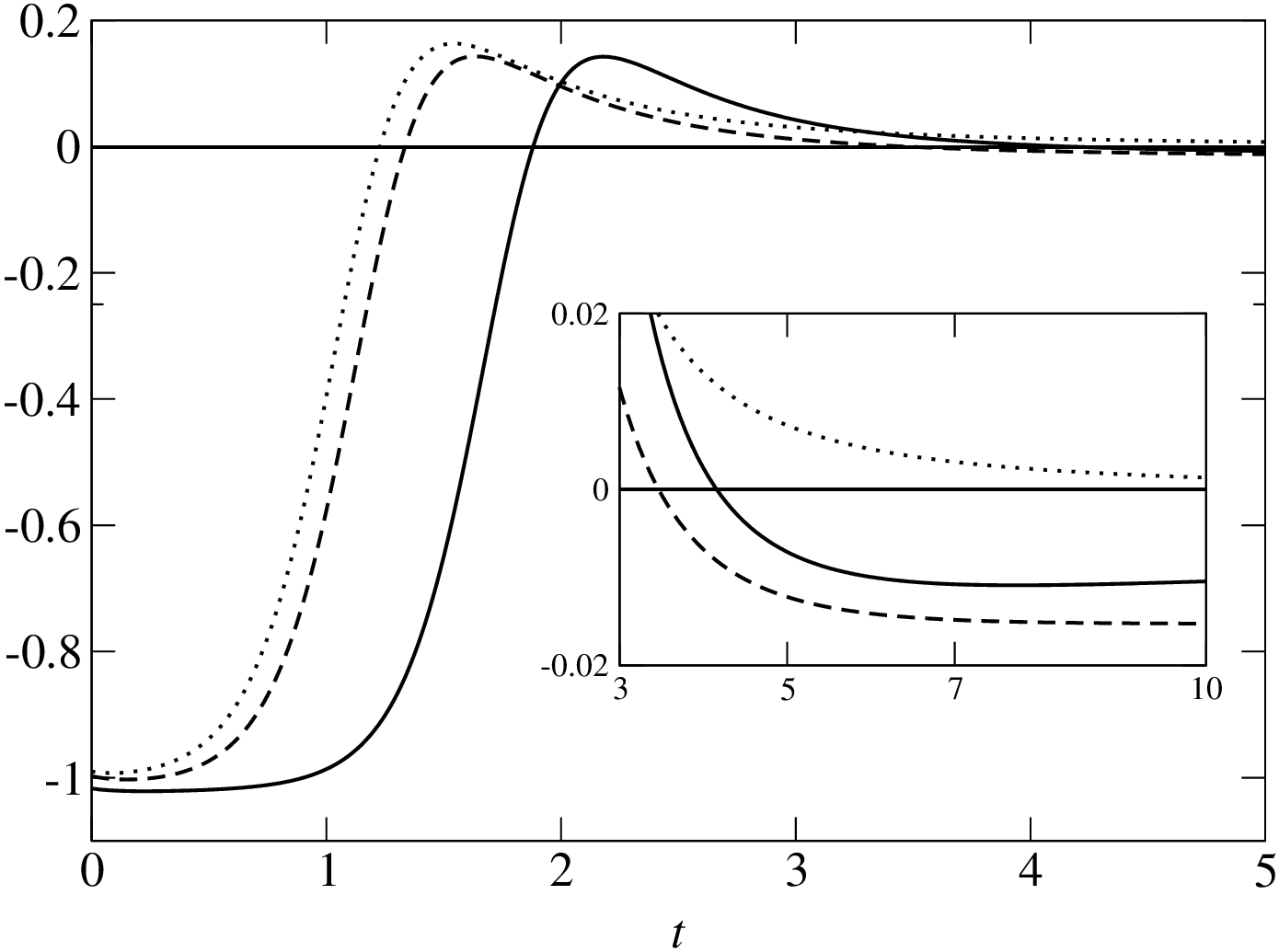}
\caption{"Total" pressure vs time for a mixture 
with quintes\-sence (straight line), for a  mixture
 with Chaplygin gas (dashed line) and for the one-component 
van der Waals fluid (dotted line). }
\end{center}\end{figure}

In the figures 3 through 5 we have chosen $w_w=0.6$, 
$\rho_X^0/ \rho_w^0=\rho_c^0/ 
\rho_w^0=0.03$, $w_X=-0.9$, $\psi=3$ and $\alpha=0.13$ and 
we proceed to discuss the results.
 
In the figure 3 it is plotted the "total" pressure 
as function of the time $t$ for a Universe considered as a: 
(a) 
mixture of a van der Waals fluid with quintessence ($p_w+p_X+\varpi$ --
straight line); (b) mixture of a van der Waals fluid
with a Chaplygin gas ($p_w+p_c+\varpi$ --
dashed line) and (c) one-component van der Waals fluid  ($p_w+\varpi$ --
dotted line). In figure 4 it was adopted  the same convention for 
the lines in the plots of
the acceleration field $\ddot a$ as a function of time $t$.
We infer from these figures that for the two models of the Universe in which
the  dark energy is present there exist: (a) an inflationary  period 
with an exponential
growth of the acceleration up to a maximum value 
where the "total" pressure remains with a constant negative value;
(b) an interval where the acceleration decreases from its maximum 
positive value 
to its maximum negative value  corresponding to a growth of the 
"total" pressure
from its maximum negative value up to a maximum positive value, and (c)
a period where a growth of the acceleration takes place from its maximum 
negative value to
a positive value and where the "total" pressure decays from its maximum 
positive
value to a negative value. Almost the same conclusions can be drawn out here
for the Universe modeled by the one-component van der Waals fluid, 
the difference being  the lack of 
a negative "total" pressure and the corresponding accelerated epoch
for the present period of the Universe, which is connected with the  
absence of the dark energy density. 
Hence, the mixtures of a van der Waals fluid with quintessence 
and with 
a Chaplygin gas can model the three periods of the Universe concerning 
its acceleration,
beginning with an inflationary accelerated period, passing through 
a past decelerated epoch and leading back to a present accelerated phase, 
while the
one-component van der Waals fluid can model only the two first periods, 
without an epoch
of present acceleration.

\begin{figure}\begin{center}
\includegraphics[width=6.4cm]{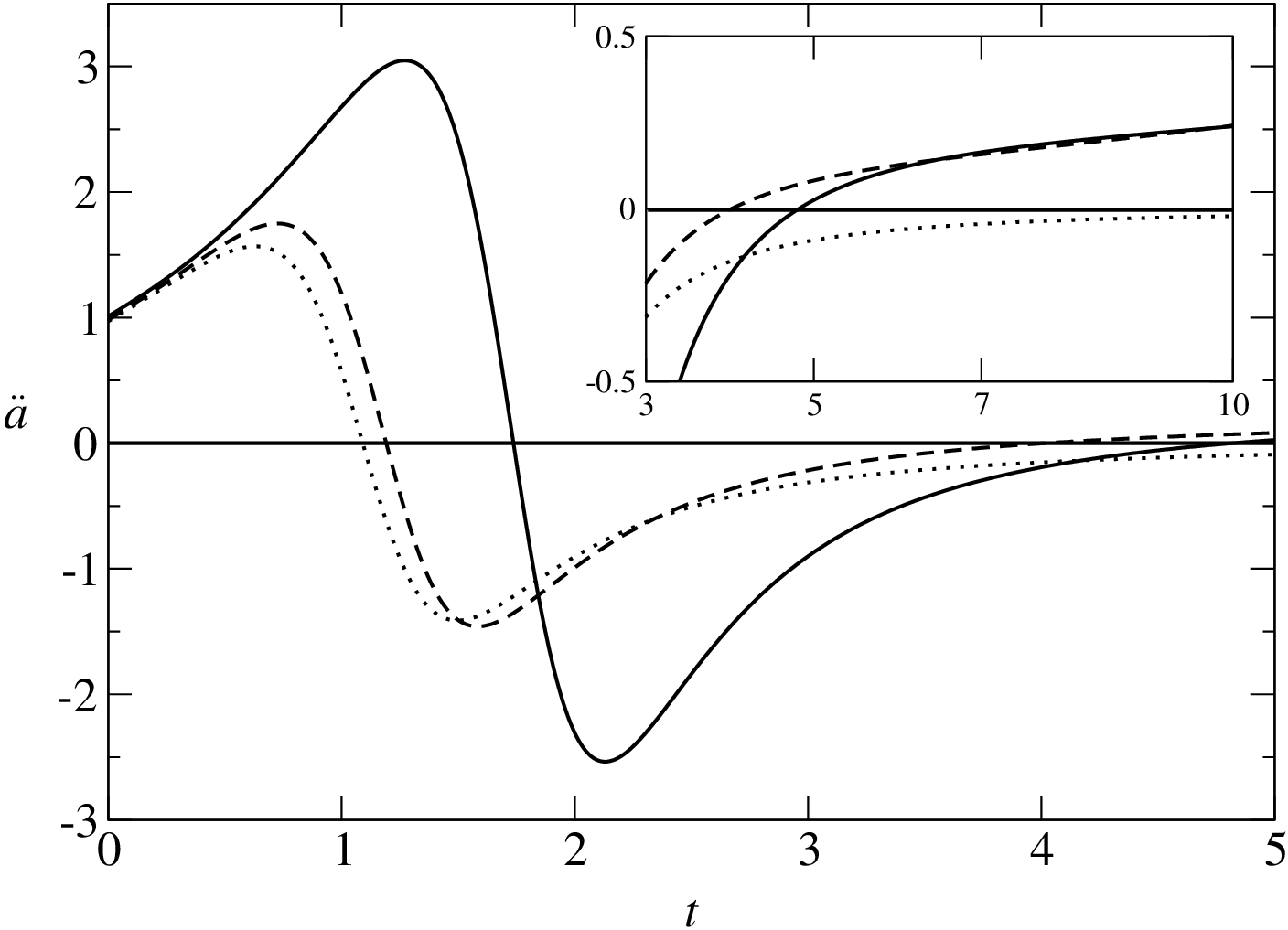}
\caption{Acceleration  vs time for a mixture 
with quintessence (straight line), for a  mixture
 with Chaplygin gas (dashed line) and for the one-component 
van der Waals fluid (dotted line).}
\end{center}\end{figure}

The energy densities of the van der Waals fluid $\rho_w$, of quintessence 
$\rho_X$ and Chaplygin gas $\rho_c$ together with the pressures of the 
quintessence $p_X$ and of the Chaplygin gas $p_c$ are plotted in figure 5
as functions of time  $t$.  As in the figures 3 and 4 the straight lines 
refer to
a mixture of the van der Waals fluid with the quintessence while the
dashed lines correspond to a mixture with the Chaplygin gas.
We conclude from figure 5 that the energy density of the van der Waals fluid 
decays more rapidly than the energy densities of the quintessence and of 
the Chaplygin
gas. This fact indicates that these two last energy densities outweigh 
the first one
at later times during the  Universe evolution, even if we consider a small 
amount 
 of the energy density of the quintessence or of the Chaplygin gas with 
respect to 
the van der 
Waals fluid at the beginning (by adjusting clocks). We infer also from this 
figure that: (a) 
the energy density of the Chaplygin gas tends to a constant value at early 
times than 
the energy density of the quintessence does, and (b) the pressure of the 
Chaplygin gas 
at later times is smaller than the pressure of the quintessence. These two 
observations
are connected with the fact that  the present acceleration  of the mixture 
with the Chaplygin gas begins at early times
than the corresponding one for the mixture with the quintessence (see also 
figure 4).

\begin{figure}\begin{center}
\includegraphics[width=6.4cm]{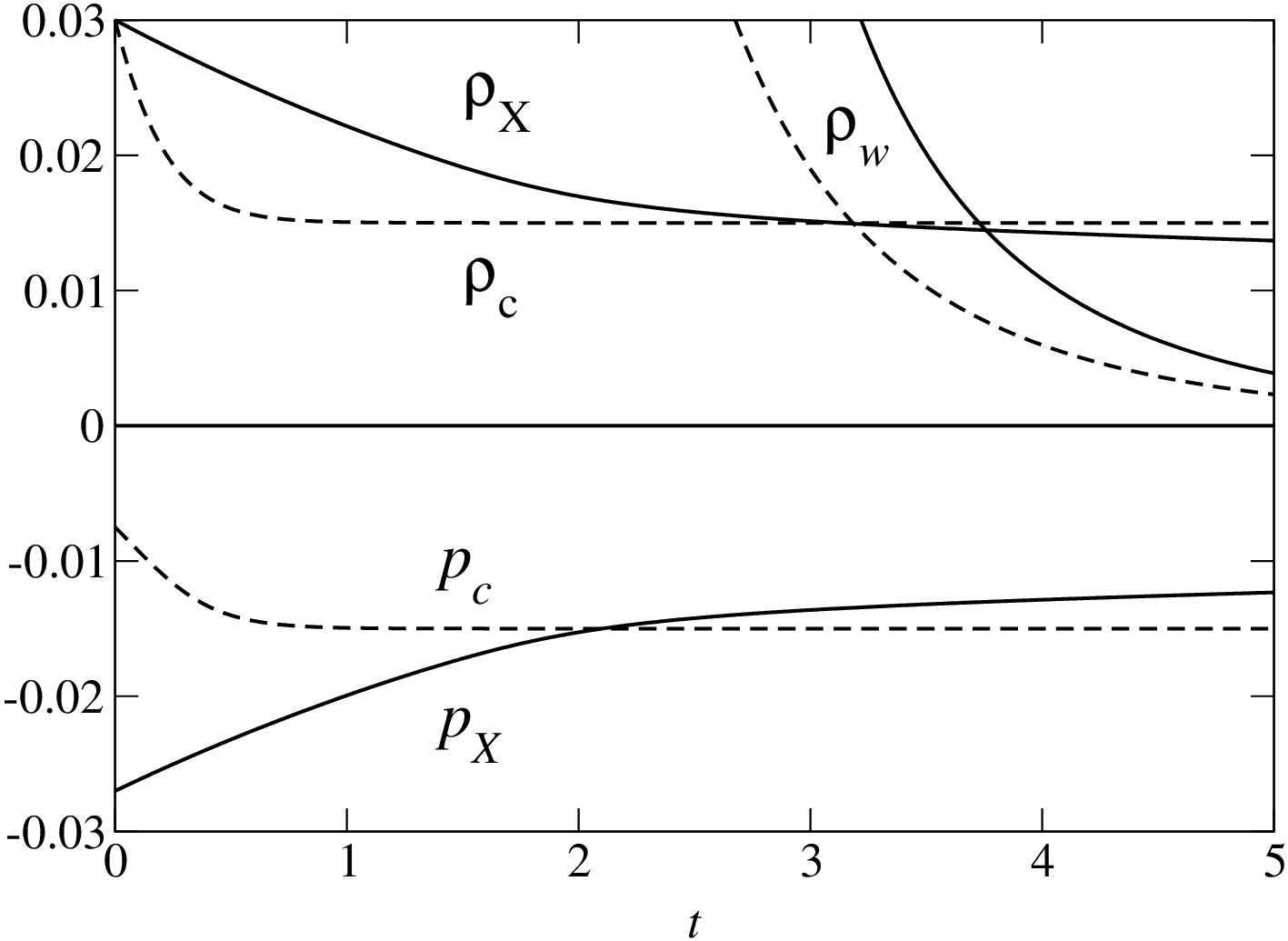}
\caption{Energy densities and pressures vs time  for a mixture 
with quintessence (straight lines) and for a  mixture
 with Chaplygin gas (dashed lines).}
\end{center}\end{figure} 

Now we shall comment on  the behavior of the solutions when the parameters 
 for the mixture of the van der Waals fluid with
the quintessence ($w_w$, $\rho_X^0/ \rho_w^0$,
$w_X$,  $\alpha$) and
 for its mixture
with the Chaplygin gas ($w_w$, $\rho_c^0/ \rho_w^0$, $\psi$, $\alpha$) are 
changed.
The period of past deceleration begins at earlier times while the period of
present acceleration begins at later times 
 when one of the following parameters changes: (a) the amount of the initial 
energy density of quintessence
with respect to the van der Waals fluid $\rho_X^0/ \rho_w^0$ decreases as 
well as 
the corresponding one for the Chaplygin gas $\rho_c^0/ \rho_w^0$,  (b) the 
strength of the
negative pressure of the quintessence $w_X$ reduces,  and (c) the amount of the
pressure-less fluid $\psi$ in the Chaplygin equation of state increases. 
Hence, all these 
parameters are connected with a more pronounced predominance of the matter 
field in
the decelerating phase of the Universe. The periods of past deceleration 
and the period of
present acceleration begins at earlier times when $w_w$ increases or $\alpha$ 
decreases.
This can be understood by recognizing that both parameters have influence 
on the 
early inflationary period, since the increase of $w_w$ leads to a less 
pronounced
inflationary period while the decrease of $\alpha$ implies that at the 
beginning
the non-equilibrium pressure has a less pronounced negative value.
 
As a final comment let us write the equation of state of the Chaplygin gas
as $p_c=w_c \rho_c$. From figure 5 one can infer that for  the times 
where the Chaplygin gas prevails over the van der Waals fluid  we
have that $-1\leq w_c<-0.8$. This last result together with the 
value $w_X=-0.9$, adopted for the quintessence, are in agreement with the 
observational constraints for $w_c$ and $w_X$ presented by some authors 
(see e.g. the works~\cite{dark}).

\end{document}